\documentclass[conference]{IEEEtran}

\usepackage[utf8]{inputenc}
\usepackage{tikz}
\usepackage{calc}
\usepackage{amsmath}
\usepackage{pgfplots}
\usepackage{amssymb}
\usepackage{listings}
\usepackage{xpatch}
\usepackage[ruled]{algorithm2e}
\usepackage{mathtools}
\usepackage{graphicx}
\usepackage{float}
\usepackage{subfig}
\usepackage[nomessages]{fp}

\usetikzlibrary{automata,positioning,arrows.meta,shapes.geometric,decorations.markings, fit}
\usetikzlibrary{calc,decorations,bayesnet}

\tikzset{>=stealth}
\usepackage[margin=1in]{geometry}
\usepackage[colorlinks,citecolor=blue,linkcolor=blue,urlcolor=blue]{hyperref}
\definecolor{light-gray}{gray}{0.7}

\usepackage[]{biblatex}
\usepackage{amsthm}

\makeatletter
\def\thmhead@plain#1#2#3{%
  \thmname{#1}\thmnumber{\@ifnotempty{#1}{ }\@upn{#2}}%
  \thmnote{ {\the\thm@headfont#3}}}
\let\thmhead\thmhead@plain
\makeatother

\newcommand*{\ARXIV}{}

\newcommand{\limitpages}[2]{
\ifdefined\ARXIV%
#2%
\else #1%
\fi%
}

\newcommand{\platformname}{World of Bugs}
\newcommand{\acr}{WOB}
\newcommand{\mlagents}{ML-Agents}

\usepackage[normalem]{ulem}

\def\BibTeX{{\rm B\kern-.05em{\sc i\kern-.025em b}\kern-.08em
    T\kern-.1667em\lower.7ex\hbox{E}\kern-.125emX}}

\pgfplotsset{compat=1.17} 
\begin{document}

\title{Learning to Identify Perceptual Bugs in 3D Video Games}

\author{\IEEEauthorblockN{1\textsuperscript{st} Benedict Wilkins}
\IEEEauthorblockA{\textit{Computer Science} \\
\textit{Royal Holloway University of London}\\
London, UK \\
   0000-0002-9107-2901}
\and
\IEEEauthorblockN{2\textsuperscript{nd} Kostas Stathis}
\IEEEauthorblockA{\textit{Computer Science} \\
\textit{Royal Holloway University of London}\\
London, UK \\ 0000-0002-9946-4037
}
}

\maketitle

\begin{abstract}
Automated Bug Detection (ABD) in video games is composed of two distinct but complementary problems: automated game exploration and bug identification. Automated game exploration has received much recent attention, spurred on by developments in fields such as reinforcement learning. The complementary problem of identifying the bugs present in a player's experience has for the most part relied on the manual specification of rules. Although it is widely recognised that many bugs of interest cannot be identified with such methods, little progress has been made in this direction. In this work we show that it is possible to identify a range of perceptual bugs using learning-based methods by making use of only the rendered game screen as seen by the player. To support our work, we have developed \platformname\ (\acr), an open platform for testing ABD methods in 3D game environments.
\end{abstract}


\begingroup

\section{Introduction}


Video games, unlike other software are intended to be an immersive experience for the user. This is achieved through narrative, deep interactivity and rich sensory presentation. This unique position in the space of software presents many development challenges, but perhaps one of the most difficult is that of testing, and in particular testing automation. Automated bug detection (ABD), the process of automatically detecting bugs through automated game exploration and bug identification, is one facet of testing automation that is gaining traction in the computer games research community. 

In industry, significant and costly steps are taken to find those bugs that can otherwise not be identified easily through automated checks. The primary methods include employing human game testers and running alpha and beta tests with real players. Even with these steps, it is common for some bugs to slip through and make it into published games.

\begin{figure}
    \centering
   
    \subfloat[\centering Environment]{\includegraphics[width=.24\textwidth]{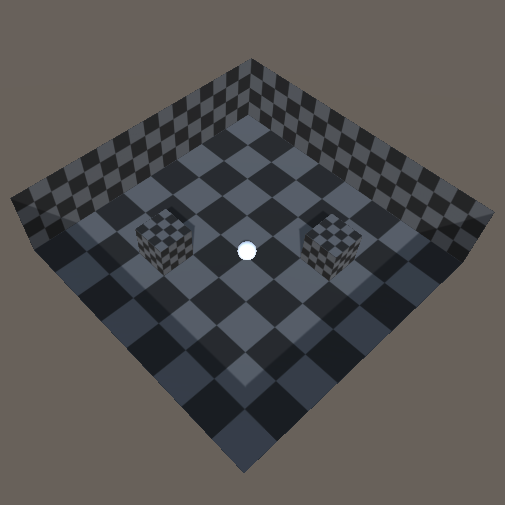}}%
    \hfill
    \subfloat[\centering
    Observation]{\includegraphics[width=.24\textwidth]{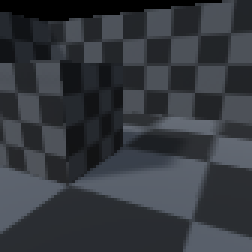}}%
    \caption{(a) shows our test environment, the agent appears as a white sphere. (b) shows the agents observation, a first-person view of the environment. }
    \label{fig:TestEnvironment}
\end{figure}

Recent attention has been on developing software agents that are able to play and explore game environments with the aim of uncovering bugs\cite{Ariyurek2021, Bergdahl2020, Gordillo2021, Gudmundsson2018}. For many works whose focus is on developing explorative agents, the bug identification problem is trivially solved for a particular class of bugs with simple rules, heuristics and exception handling. Comparatively little attention has been paid to the more complex classes of bugs that cannot be trivially identified \cite{Albaghajati2020}. 

Of the work that has been done on bug identification, the majority has focused on rule based frameworks \cite{Iftikhar2015, Varvaressos2014, Ostrowski2013}. These frameworks are generally instantiated at a higher level of abstraction, they typically analyse streams of in-game events and use logical rules to infer issues.  Practical limitations are reached when the interactions are complex and numerous. They fail to address bugs for which specifying rules is impractical, for example perceptual bugs such as rendering or audio issues. As a starting point, we are specifically focused on identifying perceptual bugs as in \cite{Wilkins2020} and \cite{Nantes2013} using methods that learn \textit{normality} from experience.

The automatic identification of complex bugs, like perceptual bugs, has not been studied in depth. We believe this is due to the problem's inherent difficulty, but also at least in part due to the difficulty of the experimental set up, as can be seen in \cite{Nantes2013}. In addition, the lack of a shared platform, or comparative baselines for testing ABD approaches has led to a somewhat fractured literature and to difficulties in evaluating and comparing performances, as can be seen on the exploration side of ABD \cite{Bergdahl2020, Gordillo2021, Prasetya2020, Shirzadehhajimahmood2021}. 

Aiming to support further research into ABD, with a focus on perceptual bugs, we have developed \platformname\ (\acr), an open platform built on top of the Unity game engine and Unity's \mlagents\ package \cite{Juliani2018}. The platform consists of a 3D first-person test environment containing a selection of specifically designed bugs, all of which manifest visually to the player in some form. Each bug can be selectively manifest, and will manifest in the same fashion as those that may be encountered during video game development \cite{Levy2009GameDE, Lewis2010, Nantes2013}.

Using data produced by our \acr\ platform we show that it is possible to identify a range of bugs using a learning-based approach, by looking only at the game screen as would be seen by a human player. These bugs include those that would otherwise be identified using rules (such as the so-called \textit{player-out-of-bounds} bug), but also extends to perceptual bugs that are otherwise difficult to identify with rules. This is an important step towards increased automation in ABD as it broadens the scope of possible bugs that can ultimately be detected. As our focus is on learning approaches to identification, in order to train our models we have generated a dataset using \acr\ consisting of a large collection of experiences. The dataset is also freely available and we hope it will serve as a baseline going forward and encourage further progress on the identification problem.

The structure of the paper is as follows. In section \ref{sec:Bug Identification} we give an in-depth discussion of bug identification and its relationship to the broader testing automation problem. In section \ref{sec:\platformname} we present the \platformname\ platform and its components. We then present our experiments and discussion in section \ref{sec:Experiments}. Finally, we present future directions and conclude in section \ref{sec:Conclusions}.

\section{Bug Identification}
\label{sec:Bug Identification}

Automating software testing is a long standing problem in software development. The problem is however notably different in video games when compared to other software \cite{Politowski2021, Santos2018, Pascarella2018}. Video games have attributes that most other software do not, they tend to present a narrative, have a rich presentation layer, and are highly interactive. Each of these attributes adds opportunity for bugs to manifest in complex ways. 

Many bugs that developers are interested in catching cannot be found via static analysis, the game must actually be played, a complex problem in itself. The process of playing and exploring a game is traditionally performed by human testers and players during the alpha or beta testing phase, but with recent developments in fields such as reinforcement learning there is potential for automation. Exploring a video game using software agents to discover bugs is a form of dynamic analysis. In software terms, this is a process of exploring and verifying paths of execution through a program, and amounts to providing various inputs to the program (fuzzing), executing it (in full), and verifying the output/program state against some criteria. In an ABD setting, the inputs are the agent's actions, and the outputs to verify are ultimately audio-visual observations. However, automatically verifying this kind of output is challenging, instead, identification methods tends to try to verify the internal state of the game.

\begin{figure}
    \centering
    \includegraphics[width=.3\textwidth]{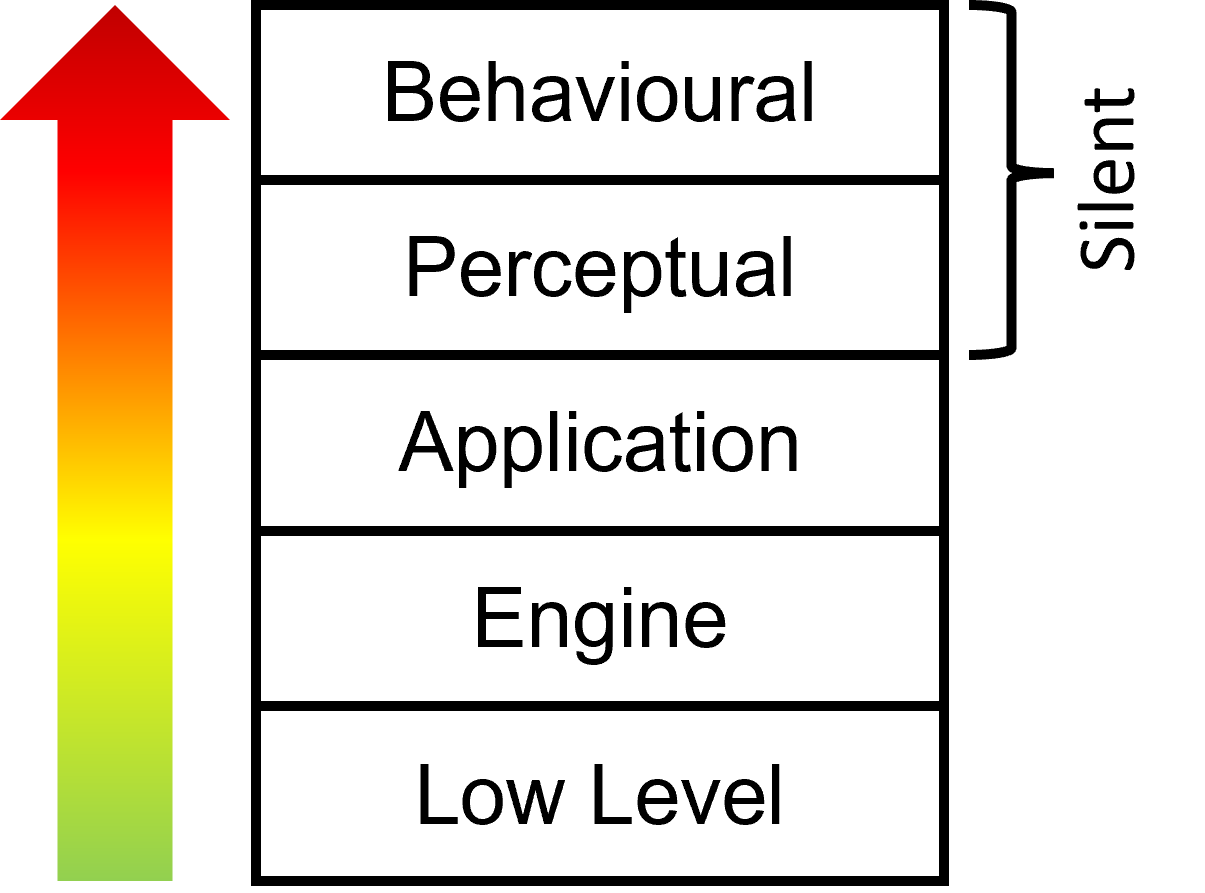}
    \caption{Hierarchy of video game bug identification showing automation difficulty. As usual, low level bugs, many internal engine and some game bugs may be identified by developer written conditions. However, perceptual and behavioural bugs require more sophisticated mechanisms for identification, the current method is to rely on human testers.}
    \label{fig:BugHierarchy}
\end{figure}

The purpose of testing is to verify that the software \textit{behaves as intended}. In our context, the manifestation of any deviation from this intended, or expected behaviour constitutes a bug. The software itself, and importantly the criteria used to verify it are a formalisation of this \textit{intention}. Both software and criteria are the same kind of object in that the criteria are also subject to miss-specification, the primary issue that developers are trying to defend against. Interestingly, the formalisation of the criteria often takes the form of code, in particular conditionals (or guards), and forms part of the software itself. This kind of formalisation is sometimes known as exception handling and is a common feature of many high-level languages. 

At a low level, a divide by zero, or segmentation fault may be caught by checking a value against some well defined bounds. The exception handling here is done at the hardware and operating system level, the error bubbles up and eventual reaches the offending program. Exception handling is also done at the game engine level and as in many software settings, at the application level. Exception handling code defines bounds or relationships on values in the internal game state. A straight forward example in video games might be to write conditions on the player's in-game position in order to catch a \textit{player-out-of-bounds} bug, a bug that allows the player to escape the playable area. A condition on the players $y$-position checked at each step in the game loop might help to uncover a so-called \textit{map-hole} bug, a type of player-out-of-bounds bug where the player falls through the level terrain. It is common practice for developers to write such checks as there many examples for which conditions can be practically specified.

Formally specifying verification criteria in this way is only practical up to a point - ultimately it is a matter of specifying bounds, or relationships between game components or properties. The internal state of a video game is often highly complex and interconnected, parts of the context may be inaccessible and checks can become prohibitively expensive. Consider a more complex instance of the \textit{player-out-of-bounds} bug in which the test should verify the player is exactly within a non-convex and possibly dynamic (in that it changes with time) level geometry; difficult, but possible with some effort. In fact, much of the work on bug identification has looked at ways to formally specify criteria, usually making use of abstractions that tend to already be present at the engine level, e.g. objects, events, activities \cite{Iftikhar2015}. Writing criteria in this way, exhaustively specifying the interactions between objects, even at an abstract level, is time consuming, and does not address the problem of limited developer imagination, it is also wholly impractical, or even impossible for certain kinds of bugs. 

Verifying properties in the internal game state or streams of events is not always enough to fully verify the final output or agents observations and this is, at least in part, the reason why human game testers are currently needed. It is useful to make a distinction between the kinds of bugs that can be easily identified without reference to the observation (by inspecting the internal game state), and those that cannot. The rendering pipeline, the program that takes the game state, including geometry, textures, lighting etc, and converts it to a visual observation (an image), may contain bugs that require inspection of the observation. These kinds of bugs, those that manifest in a visual (or auditory) manner we call perceptual bugs, see Fig. \ref{fig:BugHierarchy}. To give an example, the rendering of shadows is a procedure that relies on vast arrays of numbers (geometry, normals, light maps etc), it is wholly impractical to write criteria of the same form as for example the player position\cite{Pfau2017}. The process of verifying that shadows are rendered correctly usually amounts to manual inspection. 

In some limited settings, like for simple mobile games, it is possible to write conditions and rely on simple vision based methods \cite{Mozgovoy2018, Tuovenen2019} to identify some perceptual issues, like with positioning or with textures. To identify the more complex perceptual issues that many videos games with richer observations present, more sophisticated approaches are required. 

In addition to perceptual bugs, although not the focus of this work, there are a number of other bugs that are difficult (or impossible) to identify with the usual approach. \textit{Behavioural} is used here as an umbrella term for those bugs that require a high-level understanding of process or narrative. To give some examples, checking in-game dialogues for logical consistency, verifying the behaviour patterns of complex in game characters, or that in-game progression is possible. 

It is at this level of abstraction where human game testers tend to operate, their job is to explore the game and to identify issues that have either been overlooked by the developer (in that criteria could have be written into the code) or those for which the developer was not practically able to specify criteria. The game tester uses their knowledge of how the game is \textit{supposed} to behave (the intended behaviour), a high-level specification, and draws on their vast experience, both in game and in the real world to determine whether a particular encounter is valid or not. The question remains then, can automation be done at this level, making use of the output of the game directly as a human tester would.

Recent works on explorative agents for ABD have developed more sophisticated agents that are closer to human testers \cite{Ariyurek2021} than the more traditional rule/search based systems \cite{Shirzadehhajimahmood2021, Chang2019, Prasetya2020}. The incorporation of learning into these agents has become a key tool in allowing them to explore more effectively. We are advocating that systems that learn from experience will also be required for the identification problem, particularly if progress is to be made on identifying perceptual and behavioural bugs. Ideally, an agent would be able to leverage its experience of playing the game, along with any prior knowledge of game's intended  behaviour or potential issues to identify any bugs that manifest in its subsequent observations. Such an agent would have advantages over a traditional bug identification methods, aside from potentially being able to identify more complex perceptual or behavioural bugs, it would to some extent avoid issues like miss specification of criteria (at least by the developer/tester) and limited developer imagination. The question of how to build such an agent, along with some practical considerations is a focus of discussion in later sections.

\section{\platformname\ (\acr)}
\label{sec:\platformname}

\platformname\ is a training and test platform for automated bug detection for both exploration and identification. It has been developed in Unity and makes use of the Unity ML-Agent package \cite{Juliani2018}. The platform consists of a 3D game environment, support for developing explorative agents, and a collection of perceptual bugs that are game independent. Our hope is that progress on these bugs will be useful for the broadest audience. Perceptual bugs are also perhaps the easiest to address as they generally don't require high-level knowledge of the intended behaviour (such as knowledge about the games narrative and progression) and the application of learning methods is relatively straight forward. A full list of the implemented bugs with descriptions is available in\limitpages{the supplementary documentation}{Appendix \ref{ap:BugZoo} with further details available in the supplementary documentation.} To our knowledge \acr\ is the first freely available system that provides this functionality, and we believe it will be valuable to the ABD research community. 

We have focused our efforts on 3D first person games as a popular choice in the industry. Our test environment consists of a room, with objects situated in the room and a single agent that can move and look around. While doing so, the agent may encounter bugs manifested in its observation as they would manifest on the screen to the human player, see Fig. \ref{fig:VisualBugs}.

The platform architecture makes use of Unity's ML-Agents package, we rely on this package for executing the agent cycle and communication with Python. We wrap ML-Agent components such as sensors and agents for ease of designing agents in our context, this allows one to change agent behaviours from Python. 

An agent in \acr\ is created by default with three sensors: a main camera, which renders the view of the scene as the player would see it, a bug mask camera, which renders a mask over the scene showing in which regions there is a bug present, and a vectorized sensor which records various environment/agent properties such as position and rotation. The bug mask acts as a label for the agents observation and is instrumental in enabling machine learning to be applied to the bug identification problem. 

Bugs are defined as part of a global controller that allows them to be enabled/disabled from our Python API, newly implemented bugs can be added to the list of existing bugs in the controller. Implementing a bug requires two considerations - how to manifest the bug in the main camera and where to mask in the mask camera. The first is straight forward and often is just a matter of writing a script to modify some properties of the environment (for example deleting a texture for a missing texture bug). The second is more challenging, and for this reason we have provided some scripts/shaders to ease the process. Every bug has a tag, a unique identifier which determines its colour in the bug mask. Bug tags can be used to label bugs that are associated with particular objects in the environment, such as missing textures, corrupted geometry and bad animations. Once tagged, these objects will be rendered in the bug mask with the specific colour. Bugs such as screen tearing, flickering, and freezing, are slightly more difficult to label but typically involve a post processing step in the rendering pipeline to directly modify the final bug mask. For details on how this may be implemented refer to the supplementary documentation. The bug mask shader also contains code for rendering bugs that are not associated with a particular object and do not effect the whole scene, a good example is camera clipping (seeing through walls). It will automatically render the back-side of objects as part of the mask, that is, if the agent can see inside a geometry, the inside faces will be rendered in the mask. Additionally, it will render the sky box below a certain height, this may be used to label so-called map-hole or boundary bugs (see Fig. \ref{fig:VisualBugs}).

\renewcommand\wr{.22}
\begin{figure}%
    \centering
    \vspace{-.3cm}
    \subfloat[Texture Missing\centering]{
    \includegraphics[width=\wr\textwidth]{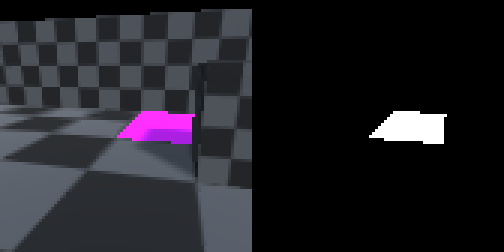}}
    \hspace{0cm} \vspace{-.3cm}
    \subfloat[Texture Corruption\centering]{
    \includegraphics[width=\wr\textwidth]{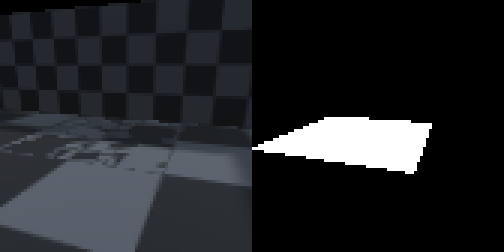}}
    \hspace{0cm} \vspace{-.3cm}
    \subfloat[Z-Fighting\centering]{
    \includegraphics[width=\wr\textwidth]{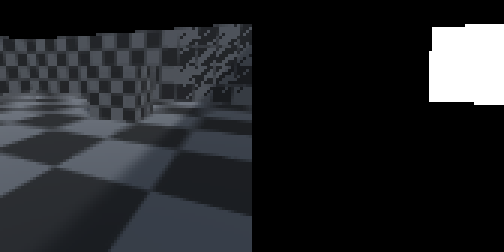}}
    \hspace{0cm} 
    \subfloat[Z-Clipping\centering]{
    \includegraphics[width=\wr\textwidth]{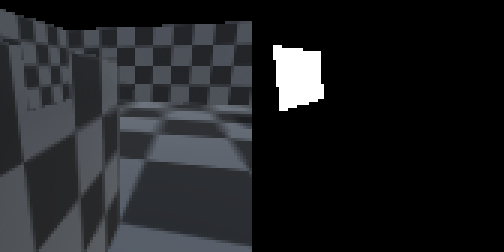}}
    \hspace{0cm} 
    \subfloat[Geometry Corruption\centering]{
    \includegraphics[width=\wr\textwidth]{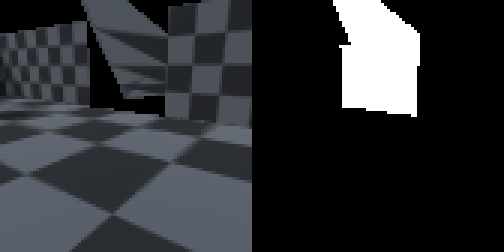}}
    \hspace{0cm} \vspace{-.3cm}
    \subfloat[Screen Tear\centering]{
    \includegraphics[width=\wr\textwidth]{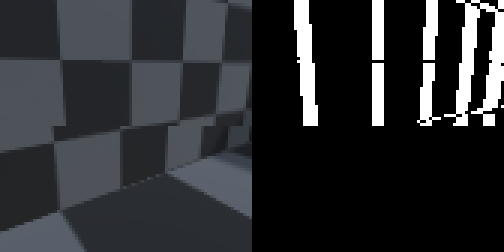}}
    \hspace{0cm} \vspace{-.3cm}
    \subfloat[Black Screen\centering]{
    \includegraphics[width=\wr\textwidth]{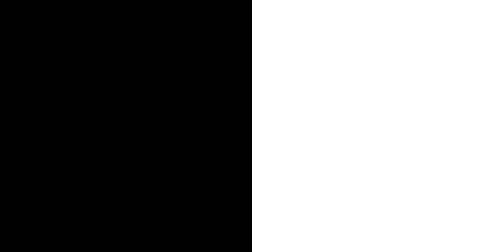}}
    \hspace{0cm} 
    \subfloat[Camera Clipping\centering]{
    \includegraphics[width=\wr\textwidth]{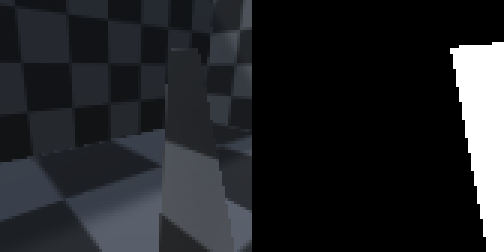}}
    \hspace{0cm} 
    \subfloat[Boundary Hole\centering]{
    \includegraphics[width=\wr\textwidth]{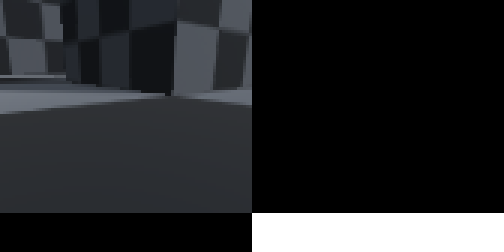}}
    \hspace{0cm}
    \subfloat[Geometry Clipping\centering]{
    \includegraphics[width=\wr\textwidth]{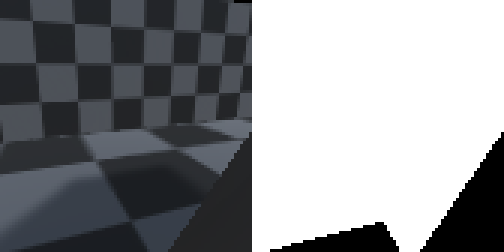}}
    \hspace{0cm}
    \caption{Bugged observations (left) with masks (right).\limitpages{see supplementary documentation for details.}{see Appendix \ref{ap:BugZoo} for details.} Bug tag colours have been removed in the masks here for readability.}
    \label{fig:VisualBugs}
\end{figure}

\subsection{Explorative Agents}
\label{sec:Explorative Agents}

\begin{figure}
    \centering
    \subfloat[1 episode (5k observations)]{\includegraphics[width=.24\textwidth]{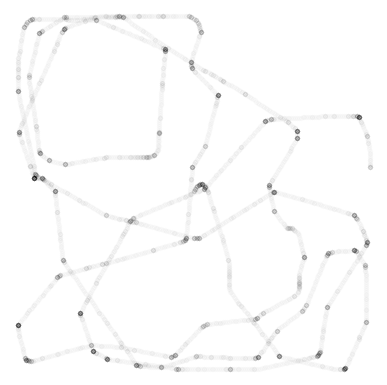}}%
    \hfill
    \subfloat[10 episodes (50k observations) ]{\includegraphics[width=.24\textwidth]{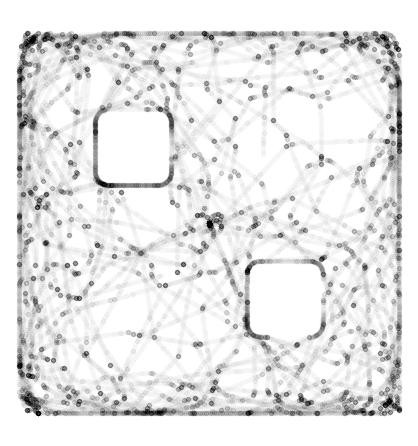}}%
    \caption{Built-in navigation agent environment coverage. }
    \label{fig:MapCoverage}
\end{figure}

Using our wrapper around the \mlagents\ package, it is possible for researchers to create and train agents to play/explore and identify bugs. This may be done via Python using our Python API which is an extension of Unity's \mlagents\ API, or from within Unity. We have developed an example agent that uses Unity's built in navigation system inspired by \cite{Prasetya2020}. This agent's behaviour is to walk around randomly while avoiding obstacles. Specifically, it selects a random point within a pre-defined front cone, refines this point so that it may be reached, computes the shortest path and moves to the closest point on this path taking the actions \texttt{forward, turn\_left, turn\_right}. At each iteration of the simulation, the agent will either choose to move to the closest point on its path, or take a random action (including \texttt{idle}) with some probability. This simple navigational agent serves, with its exploration capability, as a starting point for developing bug identification models.  
Our test environment does not have a defined goal for the agent. Although this is an important aspect of most videos games, the goal is usually not relevant for identifying the kind of bugs we are currently interested in. Omitting a goal for the agent (along with any challenging puzzles) drastically simplifies the task of developing an exploratory agent. Omitting some of the complexities of fully fledged 3D games is currently common practice when developing explorative agents \cite{Gordillo2021, Bergdahl2020} due to the difficulties that arise. For the moment we follow suit and we have opted for a simple environment to serve as a baseline, but more complex environments could be developed to test more sophisticated explorative agents. This kind of extensibility is one of the motivations for building our platform on top of a powerful game engine such as Unity.

Our built-in navigation agent is used to collect the dataset of observations that may be used to train an identification model. In a practical ABD system, both the agent and its identification model would be trained simultaneously, however having a static dataset allows reproducibility, quicker training and eases comparison of models. The dataset is described in detail in the next section. A coverage map for this agent in the test environment is shown in Fig. \ref{fig:MapCoverage}.

\subsection{Dataset}
\label{sec:Dataset}

The training data is partitioned to support both supervised and unsupervised model training. The first partition contains 300k examples (observations and actions) from 60 episodes of play which are \textit{normal} in that they contain no bugs. This is to support semi-supervised training where models are trained on only normal data and later tested on abnormal (bugged) data. The second partition contains a further 300k examples with bug masks that may be used for supervised training, or unsupervised training if the bug mask is discarded. Episodes in the second partition contain a random assortment of bugs, some episodes contain more than one kind of bug. The test set follows the second partition but consists of $\sim$ 50k examples with bug masks for each type of bug (totalling $\sim$ 500k) with each episode containing predominantly normal examples along with some bugged examples of a single given type. All observations are images are of dimension $3 \times 84 \times 84$. The dataset along with further details can be found \href{https://www.kaggle.com/benedictwilkinsai/world-of-bugs}{here}\footnote{https://www.kaggle.com/benedictwilkinsai/world-of-bugs}.
\section{Experiments}
\label{sec:Experiments}

In our experiments the focus is on learning-based methods for bug identification. Our platform is tailored to these methods (although rule-base methods are not excluded). The aim of our experiments is to demonstrate that learning-based methods are a viable alternative to the more traditional rule-based methods for identifying video game bugs. In addition, to show that they may be used as kind of \textit{catch-all}, to identify both perceptual bugs and those that can otherwise be identified with rule-based approaches, such as the \textit{player-out-of-bounds} bug.

During the video game development process, at which time we might want to make use of an ABD system, it is unlikely that one has access to a bugless version of the game. This places us in an unsupervised regime, as we do not have access to labelled data. The bug identification problem becomes vastly more challenging in this regime. There are however methodologies that aim to deal (at least partially) with this challenge by re-framing the fully unsupervised problem as iterative and semi-supervised. The problem in its essence is transformed into a problem of finding differences in what is observed in successive versions of the game. Differences may be desirable (features or bug fixes) or undesirable (bugs) according to a human in the loop. For our purposes as a proof of concept, our models are trained on a single \textit{normal} version of the game, and tested later on different version each containing one type of bug.

In practical terms, a human developer will need to interact with an ABD system in order to resolve any issues that have been found. The goal of developing such a system in the first place is to reduce the manual workload required for testing. The workload is drastically reduced by having an explorative agent play the game, but in order to identify bugs, the ABD system will need to produce a report, just like a human tester would. The simplest kind of report would be a list of observations, sorted by the score the model has given them. With workload reduction in mind, this kind of report has some implications for our preferences on the systems capabilities. A bug identification model should always identify a bug when encountered by the explorative agent, it should have a high true-positive rate (TP). It is also important that the model does not identify observations as bugged when there are not, that is, it should have a low false-positive rate (FP), otherwise the developer will need to sift through many observations that are irrelevant to the task at hand. Model precision ($TP / TP + FP$) is a useful measure of model performance in this setting. 
Many bugs will manifest in a multitude of ways at different times during the agents experience. The bug identification model needs only to identify one of these instances for the developer to make some progress towards a fix. This should indicate a lesser importance to the false-negative rate (FN) of the model, as long as the developer has some understanding of the model's limitations with regard to different classes of bugs. The results presented reflect these considerations, with a focus on model precision. 

\subsection{Reconstruction}

Auto-encoders are commonly used to detect anomalies in visual scenes\cite{Yang2021, Xu2019, Ribeiro2018}. They work by projecting input to a low dimensional space using an encoder network, and then reconstructing it using a decoder network. The error in the reconstruction when compared to the original input may be used as a score for how \textit{anomalous} a particular input is, a larger error tends to positively correlate with anomalousness. The reconstruction approach only works if the auto-encoder rarely (or never) encounters anomalous examples during training, the training is in this sense semi-supervised. The auto-encoder used in our experiments is a convolutional network. We use Structural Similarity Index (SSIM) \cite{Zhou2004} with a small MSE penalty term as the objective function, and squared L2 norm as the anomaly score. We found that the reconstructions using SSIM + MSE penalty were visibly more convincing and the resulting models performed better than those trained with Mean Squared Error (MSE) or with Binary Cross Entropy loss (BCE) or SSIM alone. Results can be found in Fig. \ref{fig:curves}. Training details including model architectures are given in\limitpages{the supplementary documentation}{Appendix \ref{ap:training-details}}, all code is available \href{https://github.com/BenedictWilkins/world-of-bugs-experiments}{here}\footnote{https://github.com/BenedictWilkins/world-of-bugs}.

In some cases, the bug mask contains only a few pixels, in these cases there may not be enough information for a model to determine whether the particular observation is bugged. For this reason we present results for different pixel thresholds $\tau$ in the bug masks.  (i.e. an observation is labelled as a containing a bug if the bug mask contains more than $\tau$ coloured pixels).

\limitpages{
\limitpages{
\begin{figure*}
    \centering
    \includegraphics[width=\textwidth]{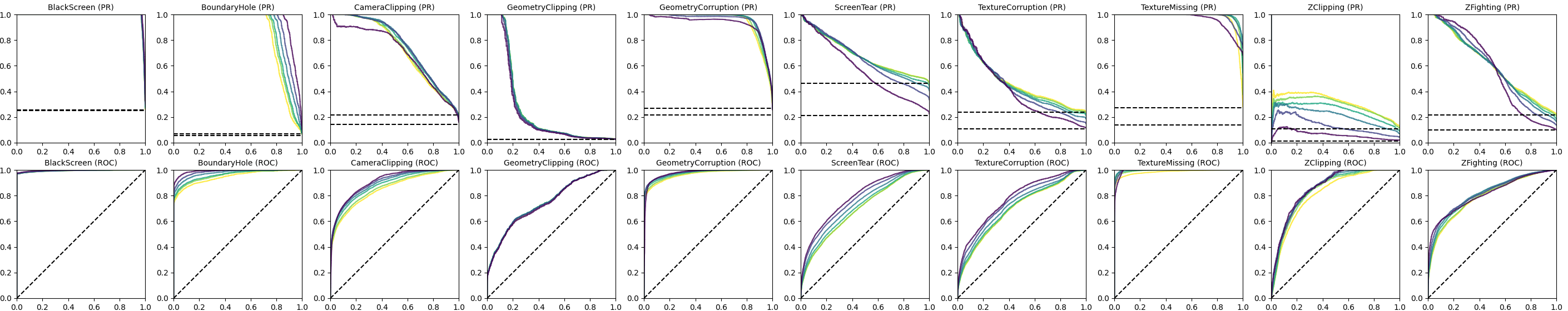}
    \caption{Precision Recall Curves (top) and Receiver Operate Characteristic Curves (bottom) for each kind of bug we tested: Black Screen, Boundary Hole, Camera Clipping, Geometry Clipping, Geometry Corruption, Screen Tear, Texture Corruption, Texture Missing, Z-Clipping, Z-Fighting (left-right). Dotted lines show random performance for $\tau=0$ (light/yellow) and $\tau=400$ (dark/purple). AUC scores are reported in Table \ref{tab:AUC}. }
    \label{fig:curves}
\end{figure*}
\begin{table*}[]
    \centering
    \begin{tabular}{|l|} \hline 
                                \\ \hline
        Label Threshold $\tau$ \\ \hline
        BlackScreen          \\
        BoundaryHole         \\
        CameraClipping       \\
        GeometryClipping     \\
        GeometryCorruption   \\
        ScreenTear           \\
        TextureCorruption    \\
        TextureMissing       \\
        ZClipping            \\
        ZFighting            \\ \hline

    \end{tabular}
     \begin{tabular}{| c | c | c | c | c | c | } \hline
        \multicolumn{6}{| c |}{Precision Recall AUC} \\ \hline
            0 & 10 & 50 & 100 & 200 & 400 \\ \hline
            0.990 & 0.990 & 0.990 & 0.990 & 0.991 & 0.992 \\
            0.939 & 0.951 & 0.956 & 0.958 & 0.960 & 0.965 \\
            0.748 & 0.756 & 0.772 & 0.782 & 0.780 & 0.743 \\
            0.156 & 0.157 & 0.155 & 0.153 & 0.151 & 0.142 \\
            0.934 & 0.940 & 0.947 & 0.951 & 0.952 & 0.940 \\
            0.676 & 0.675 & 0.670 & 0.664 & 0.641 & 0.562 \\
            0.471 & 0.465 & 0.455 & 0.449 & 0.432 & 0.403 \\
            0.973 & 0.986 & 0.989 & 0.990 & 0.981 & 0.962 \\
            0.319 & 0.300 & 0.259 & 0.214 & 0.143 & 0.072 \\
            0.600 & 0.598 & 0.596 & 0.593 & 0.590 & 0.579 \\ \hline
    \end{tabular} 
    \begin{tabular}{| c | c | c | c | c | c | } \hline
        \multicolumn{6}{| c |}{Receiver Operate Characteristic AUC} \\ \hline
            0 & 10 & 50 & 100 & 200 & 400 \\ \hline
            0.993 & 0.994 & 0.994 & 0.994 & 0.994 & 0.996 \\
            0.981 & 0.986 & 0.987 & 0.988 & 0.989 & 0.990 \\
            0.853 & 0.864 & 0.883 & 0.896 & 0.905 & 0.911 \\
            0.742 & 0.745 & 0.749 & 0.756 & 0.758 & 0.743 \\
            0.957 & 0.963 & 0.968 & 0.972 & 0.975 & 0.976 \\
            0.692 & 0.694 & 0.709 & 0.735 & 0.765 & 0.788 \\
            0.698 & 0.702 & 0.709 & 0.730 & 0.759 & 0.781 \\
            0.983 & 0.993 & 0.996 & 0.996 & 0.995 & 0.992 \\
            0.812 & 0.837 & 0.848 & 0.854 & 0.862 & 0.864 \\
            0.789 & 0.794 & 0.810 & 0.819 & 0.821 & 0.814 \\\hline
    \end{tabular}
   
    \caption{Receiver Operate Characteristic and Precision Recall AUC scores for varying thresholds $\tau$.}
    \label{tab:AUC}
\end{table*}
}{
\begin{figure*}
    \centering
    \includegraphics[width=\textwidth]{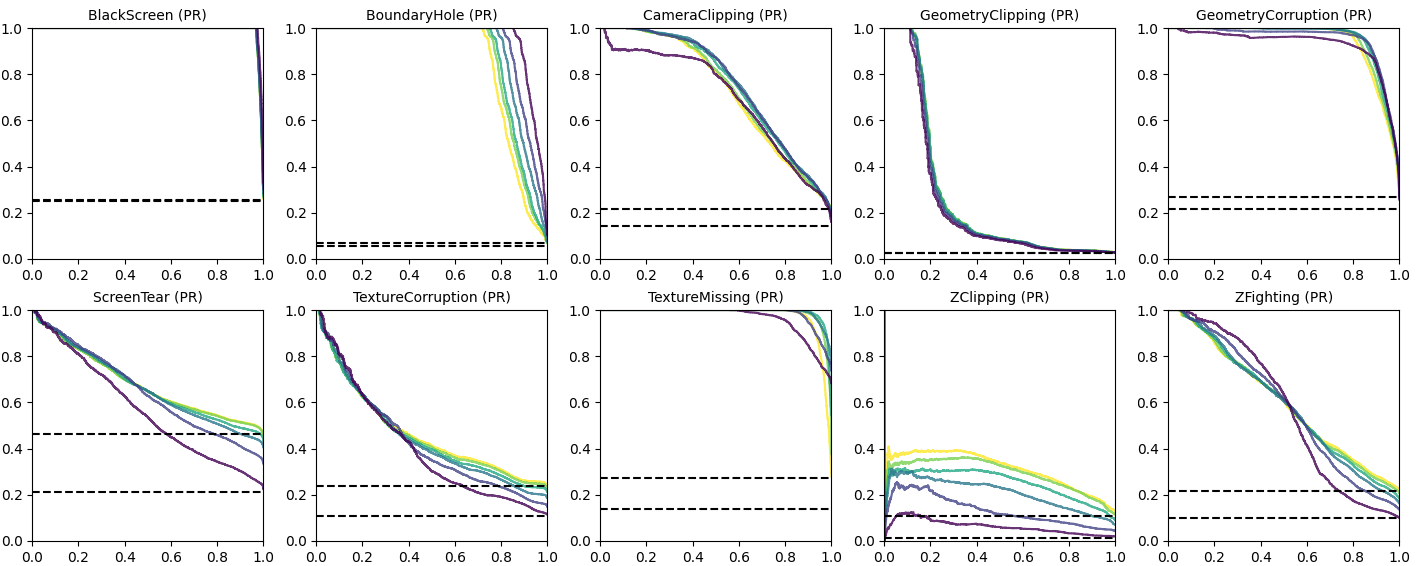}
    \includegraphics[width=\textwidth]{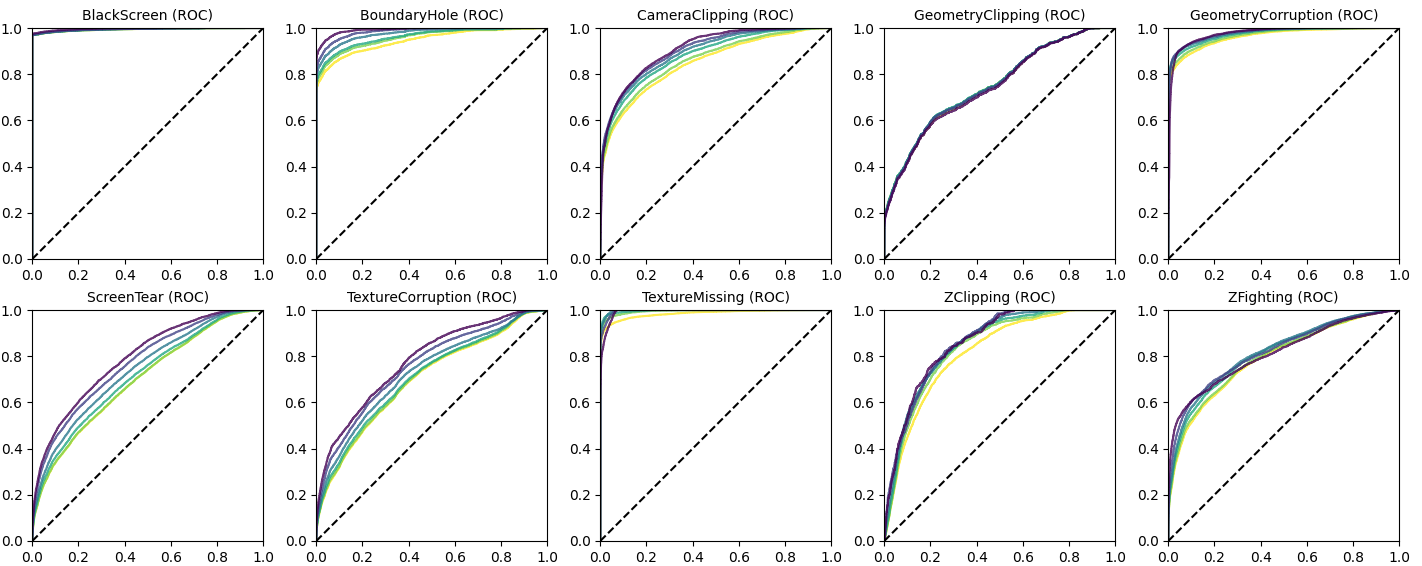}
    \caption{Precision Recall Curves and Receiver Operate Characteristic Curves for each kind of bug we tested. Dotted lines show random performance for $\tau=0$ (light/yellow) and $\tau=400$ (dark/purple). AUC scores are reported in Table \ref{tab:AUC}. }
    \label{fig:curves}
\end{figure*}

}}{}
\subsection{Discussion}
\label{sec:Discussion}

The reconstruction-based approach was able to identify many of the bugs implemented in \acr, see Fig. \ref{fig:curves}. It was particularly successful in identifying missing textures, boundary holes, camera clipping and corrupted geometry, as well as the significantly easier black screen bug. But was less good at identifying some of the more subtle bugs, those that did not have a large impact on the observation or that look similar to normal experience. This is perhaps best highlighted by the geometry clipping bug, a bug where the player is able to walk through objects. This bug is very similar to camera clipping where performance was better. The difference being that the model was not able to identify cases where it was fully enclosed in the geometry of the object. The poorer performance could be because the back-side of the object geometry is not rendered. The observation looks normal as the object simply disappears from view leaving a normal background scene, but it is still labelled as bugged as the back-sides are rendered in the bug mask. A model that learns from sequences of observations may perform better in this case. The result obtained for the camera clipping bug suggests that even this simple approach would be helpful in practice for detecting geometry clipping. After all, only the point at which the player passes through the object boundary needs to be identified for a developer/tester to be made aware of the problem. They can then infer the underlying issue by manually inspecting the observations to follow (i.e. to determine if it is a geometry clipping, camera clipping issue or otherwise).


\section{Conclusions}
\label{sec:Conclusions}



The \platformname\ platform presented in this work provides a starting point for researchers and practitioners to address the ABD problem more broadly. We have demonstrated its use with a simple explorative agent and learning-based approach to identifying a range of perceptual bugs. Our experiments have shown that models that learn what is \textit{normal} from the agent's experience are able to identify some perceptual bugs. Importantly, we showed that they are able to identify perceptual issues that are otherwise hard to identify with rules-based approaches. This result is encouraging as the models we tested are relatively simple and do not fully leverage recent advancements in visual and time-series anomaly detection. We expect that with more sophisticated learning approaches and with the right inductive biases further progress could be made on perceptual bugs in environments visually more complex than our test environment. 

Aside from exploring more complex visuals, game play mechanics, perceptual modes (e.g. audio) and other kinds of bugs beyond those presented in this work, there are other directions that require further study, many of which are supported by our platform. For example, integrating exploration and identification in a meaningful way, to direct exploration towards regions where issues may be more common, and improving issue reports, perhaps through bug segmentation or clustering.




\endgroup

\printbibliography
\clearpage

\limitpages{
}{

\appendix


\subsection{Training Details \& Model Architecture}
\label{ap:training-details}
We trained a large number of models with varying architectures and hyper parameters to attain the results presented in this work. Our focus was on relatively simple auto-encoder architectures, all of the architectures/parameters along with further details and trained models can be found \href{https://github.com/BenedictWilkins/world-of-bugs-experiments}{here}\footnote{https://github.com/BenedictWilkins/world-of-bugs-experiments}. The model for which we present results was trained for 5 epochs on the full 300k normal training examples and tested on the full test dataset (500k) examples. Training took approximately 40 minutes on a single NVIDIA Tesla M60 on a 32 core machine.

We used the well known Adam optimiser with a constant learning rate of $0.0005$, no weight decay, $\beta=(0.9,0.999)$. We used a combination of SSIM and MSE weighted with $0.9$ and $0.1$ respectively as the objective function. Although SSIM produces better reconstructions we found it was subject to some instability which was reduced by introducing a small MSE penalty term. The scoring function used to rank observations is squared L2 between the observation and reconstruction. The model architecture used is presented in Table \ref{tab:model-architecture}. 

\begin{table*}
    \centering
    \begin{tabular}{|l|l|l|} \hline
        Layer Type & Activation & Parameters \\ \hline
        \multicolumn{3}{| l |}{Encoder Network, input\_dim=(3,84,84)} \\ \hline
        Conv2D & LeakyReLU & in\_channel=3, out\_channel=16, kernel\_size=6, stride=1 \\ 
        Conv2D & LeakyReLU & in\_channel=16, out\_channel=32, kernel\_size=5, stride=2 \\ 
        Conv2D & LeakyReLU & in\_channel=32, out\_channel=64, kernel\_size=6, stride=1 \\ 
        Conv2D & LeakyReLU & in\_channel=64, out\_channel=128, kernel\_size=5, stride=2 \\ 
        Conv2D & LeakyReLU & in\_channel=128, out\_channel=128, kernel\_size=5, stride=2 \\ 
        Conv2D &           & in\_channel=128, out\_channel=128,kernel\_size=5,stride=1 \\ \hline
        \multicolumn{3}{| l |}{Decoder Network, input\_dim=(2,2,128)=(512,)} \\ \hline
        ConvTransposed2D & LeakyReLU & in\_channel=128, out\_channel=128, kernel\_size=5, stride=1 \\ 
        ConvTransposed2D & LeakyReLU & in\_channel=128, out\_channel=128, kernel\_size=5, stride=2 \\
        ConvTransposed2D & LeakyReLU & in\_channel=128, out\_channel=64, kernel\_size=5, stride=2 \\ 
        ConvTransposed2D & LeakyReLU & in\_channel=64,  out\_channel=32, kernel\_size=6, stride=1 \\ 
        ConvTransposed2D & LeakyReLU & in\_channel=32,  out\_channel=16, kernel\_size=5, stride=2 \\ 
        ConvTransposed2D &           & in\_channel=16,out\_channel=3,kernel\_size=6,stride=1 \\ \hline
    \end{tabular}
    \caption{Model Architecture which we present results for. Total trainable parameters = 2,225,379 }
    \label{tab:model-architecture}
\end{table*}

\subsection{Bug Zoo}
\label{ap:BugZoo}

Below is a list of the bugs implemented in our platform along with a brief description of each. All of the bugs presented are encountered by developers in real development scenarios \cite{Levy2009GameDE, Lewis2010}. Note that these bugs are game independent in the sense that the associated scripts/shaders/game components may be used in other Unity3D projects allowing researchers/developers to train detection models in environments more complex than our test environment. 

\subsubsection{Camera Clipping}
For efficiency reasons, rendering pipelines cull geometry that is out of view. Cameras are a useful abstraction that are used by many engines to describe what can be seen on screen, the cameras view frustum is used to cull unseen geometry. If this frustum is too small, some geometry that should be rendered is not. In the example below, the near clipping plane has been set too far away from the player. At certain view points, when close to an object some of its geometry is culled leading to the “see through” effect seen below. In some circumstances this bug allows players to see into areas they shouldn't be able to (looking through walls/floors). This bug may only manifest itself with certain objects, particularly ones that have an odd shape (non-convex or pointy) and so it may not be immediately obvious to a developer that there is an issue. \textit{Implementation:} Modify the near clipping plane.

\subsubsection{Texture Corruption}
A texture may render incorrectly/become corrupt for various reasons, for example, when texture offsets incorrectly set, the UV map is incorrect or as a result of shader related issues. Similar effects may also be due to lighting problems, for example if the vertex normals are incorrectly set.  \textit{Implementation:} the UV map/offsets are modified to corrupt the texture. 

\subsubsection{Texture Missing}
When a texture is missing from an object, which may happen if a developer forgets to set the texture, its missing/inaccessible in the file system or otherwise, the shader responsible for rendering the texture will typically resort to rendering the object it in one colour (typically bright pink) or resort to a default texture. \textit{Implementation:} remove the texture from a chosen game object.

\subsubsection{Z-Clipping}
Rendering pipelines often have layers, layers contain geometry to be rendered and are queued and rendered in order. Pixels in the layers rendered subsequently are replaced or are blended with pixels in previous layers. If an object geometry is placed in the wrong layer, it may be rendered on top of others leading to some strange spatial effects (which may or may not be desirable). \textit{Implementation:} places a chosen game object in a layer that is always rendered last.

\subsubsection{Z-Fighting}
Z-Fighting happens when two surfaces have the same depth (z). The renderer/shader does not know which to show first and this results in a mixing of textures from the two surfaces. A flickering effect may also occur during when the players view shifts. \\ \textit{Implementation:} copies a game object, modifies its texture and places it exactly at the position of the original. 

\subsubsection{Geometry Corruption}
Similarly to Texture Corruption, object geometry can also become corrupted. A geometry is corrupted when some of its vertices are incorrectly placed relative to the other vertices. They tend to happen during animations that depend on physics with flexible/dynamic geometries, but can also happen with static geometry (although this is less common). 
\textit{Implementation:} randomly modifies the vertex positions of a chosen game object geometry over time. 

\subsubsection{Black Screen}
A black screen usually happens when there is an issue with the rendering pipeline, leading to a screen that is black (nothing is rendered). \textit{Implementation:} Render a black texture to the screen in a post processing step. 

\subsubsection{Screen Tear}
Screen tearing happens when a monitors refresh rate doesn't match the GPU frame rate. This leads to information multiple frames being rendered to the screen as a single frame, leading to a tearing effect. Although typically a hardware problem, it represents an interesting and often subtle visual issue. \textit{Implementation:} Record previous frames and render them embedded in the current frame. 
 
\subsubsection{Geometry Clipping}

A bug that manifests in a visually similar way to camera clipping, but has a different cause. If collisions between the player and an object are not computed correctly, the player may be able walk inside the object and so is able to see inside it. 
\textit{Implementation:} Disable the collision box of an object. Note: similar collision based bugs between two (non-player) objects, or \textit{ghost objects} also manifest visually, this has not been explored in our work but would be an interesting case to study.

\subsubsection{Boundary Hole / Player out of Bounds}

A specific kind of Geometry Clipping bug in which the player falls through the terrain due to gravity, or is able to walk out of the bounds of play.\textit{Implementation:} Disable the collision box of a boundary object/terrain.

}

\end{document}